\documentclass[twocolumn, floatfix,showpacs,preprintnumbers,amsmath,amssymb,pra,superscriptaddress,longbibliography]{revtex4-2}
\usepackage{color}
\usepackage[usenames,dvipsnames,svgnames,table]{xcolor}
\usepackage[colorlinks=true,linkcolor=blue,urlcolor=blue,citecolor=blue]{hyperref}
\usepackage{mathtools}
\usepackage{graphicx}
\usepackage{dcolumn}
\usepackage{array}
\usepackage{lipsum}
\usepackage{bm}
\usepackage{subfigure}
\usepackage{amssymb}
\usepackage{multirow}
\usepackage{tabularx}
\usepackage{amsmath}
\usepackage{braket}
\usepackage{csquotes}
\usepackage{enumitem}
\graphicspath{{plots/}}
 \usepackage{lipsum}
\usepackage{mathrsfs}
\usepackage{MnSymbol}

\usepackage{algorithm}
\usepackage{algpseudocode}
\usepackage{todonotes}

\usepackage[normalem]{ulem} 

\begin{document}
\title{Temperature dependence of the dynamic structure factor of the
electron liquid via analytic continuation}

\author{Thomas Chuna}
\email{t.chuna@hzdr.de}
\affiliation{Center for Advanced Systems Understanding (CASUS), Helmholtz-Zentrum Dresden-Rossendorf (HZDR), D-02826 G\"orlitz, Germany}

\author{Maximilian B\"ohme}
\affiliation{Quantum Simulations Group, Physics and Life Science Directorate, Lawrence Livermore National Laboratory (LLNL), California 94550 Livermore, USA}

\author{Tobias Dornheim}
\email{t.dornheim@hzdr.de}
\affiliation{Institute of Radiation Physics, Helmholtz-Zentrum Dresden-Rossendorf (HZDR), D-01328 Dresden, Germany}
\affiliation{Center for Advanced Systems Understanding (CASUS), Helmholtz-Zentrum Dresden-Rossendorf (HZDR), D-02826 G\"orlitz, Germany}

\begin{abstract}We present new analytic continuation results for the dynamic structure factor $S(\mathbf{q},\omega)$ of the uniform electron liquid based on quasi-exact \emph{ab initio} path integral Monte Carlo (PIMC) data for the imaginary-time density--density correlation function $F(\mathbf{q},\tau)$ across a broad range of temperatures.
For this purpose, we employ both a traditional maximum entropy method solver, and a pre-optimized sparse Gaussian kernel representation as it has been implemented in the recent \texttt{PyLIT} package [Benedix Robles \textit{et al.}, \textit{Comp.~Phys.~Comm.}~\textbf{319}, 109904 (2026)], and we identify potential advantages and disadvantages in both.
We expect our results to be interesting for a broad range of topics, including the interpretation of x-ray Thomson scattering experiments with extreme states of matter and the construction of improved exchange--correlation kernels for linear-response time-dependent density functional theory.
\end{abstract}

\maketitle



\section{Introduction}

Understanding the dynamic properties of non-ideal (i.e., interacting) quantum many-body systems is of key importance for a great variety of research field in physics, quantum chemistry, material science and related disciplines~\cite{giuliani2008quantum,bonitz_book,pines}.
Of particular importance are the dynamics of the electrons, which are often described using different time-dependent density functional theory (TD-DFT) methodologies~\cite{book_Ullrich,Moldabekov_MRE_2025,kononov2025realtimetimedependentdensityfunctional,French_PRB_2010,dynamic2,White_ElectronicStructure_2025}.
More specifically, real-time TD-DFT~\cite{dynamic2,kononov2025realtimetimedependentdensityfunctional} is, in principle, capable of estimating non-linear and non-equilibrium effects as they are relevant, e.g., for stopping power calculations~\cite{kononov2025realtimetimedependentdensityfunctional,kononov2025nonlineareffectslightionstopping}. Yet, calculations require as input a dynamic exchange--correlation (XC) potential, which has to be approximated in practice.
Alternatively, the linear-response TD-DFT approach~\cite{Gross_PRL_1985,Moldabekov_MRE_2025,Moldabekov_PRR_2023}, which is, by construction, limited to the estimation of the equilibrium dynamics of the simulated system, requires as input the usual static XC-functional and, additionally, the dynamic XC-kernel $K_\textnormal{XC}(\mathbf{q},\omega)$, which is related to the former by a double functional derivative.

The real-time formalism discussed above is supported by the \emph{imaginary-time} formalism~\cite{Dornheim_MRE_2023,Dornheim_JCP_ITCF_2021}. Many XC-functionals for DFT~\cite{Perdew_Zunger_PRB_1981,vwn,Perdew_PRL_1996,ksdt,groth_prl,Karasiev_PRL_2018} and and XC-kernels for TD-DFT~\cite{Dabrowski_PRB_1986,cdop,Constantin_PRB_2007,Panholzer_PRL_2018,dornheim_ML,Adrienn_PRB_2020,Ramakrishna_PRB_2021}) are informed by properties of the uniform electron gas (UEG), which are described with very high accuracy by \emph{ab initio} quantum Monte Carlo (QMC) simulations, for the ground state~\cite{Ceperley_Alder_PRL_1980,Ortiz_PRL_1999,Spink_PRB_2013,moroni2} and at finite temperatures~\cite{review,Brown_PRL_2013,dornheim_prl,dornheim_ML,Dornheim_PRL_2020,Hou_PRB_2022,Dornheim_PRB_2025}. Unfortunately, obtaining estimates of dynamic properties is particularly challenging because widely used path integral Monte Carlo (PIMC)~\cite{cep} simulations can only access dynamic information in the form of various imaginary-time correlation functions (ITCFs)~\cite{Boninsegni1996,Rabani_PNAS_2002,boninsegni1,Filinov_PRA_2012,efremkin2026computationthermalconductivitybased}, which, in turn, are related to the spectral properties of interest via an integral transform of the form~\cite{Jarrell_PhysRep_1996}
\begin{eqnarray}\label{eq:kernel}
    F(\tau) = \int_{-\infty}^\infty \textnormal{d}\omega\ K(\omega,\tau)\ S(\omega)\ ,
\end{eqnarray}
with $K(\omega,\tau)$ being the analytically known integration kernel.
The task at hand is thus to numerically invert Eq.~(\ref{eq:kernel}) to reconstruct the spectrum $S(\omega)$ from the PIMC results for $F(\tau)$.
This \emph{analytic continuation} is a notoriously difficult and, in fact, ill-posed problem~\cite{Jarrell_PhysRep_1996,Goulko_PRB_2017,Chuna_JPA_2025,chuna2025noiselesslimitimprovedpriorlimit}; even small Monte Carlo error bars on $F(\tau)$ can lead to substantially different results for $S(\omega)$.

Nevertheless, the pressing need for an accurate description of dynamic properties has led to a remarkable number of developments in the field of analytic continuation, see, e.g., Refs.~\cite{MEM_original,Jarrell_PhysRep_1996,Mishchenko_PRB_2000,Vitali_PRB_2010,Sandvik_PRE_2016,Bertaina_GIFT_2017,dornheim_dynamic,dynamic_folgepaper,Otsuki_PRE_2017,Otsuki_JPSJ_2020,SHAO20231,Filinov_PRB_2023,Chuna_JPA_2025,BENEDIXROBLES2026109904}.
To the best of our knowledge, the first analytic continuation results for the dynamic structure factor $S(\mathbf{q},\omega)$ of the UEG---the key observable in x-ray Thomson scattering (XRTS) experiments with matter under extreme conditions and beyond~\cite{siegfried_review,Dornheim_review,gregori2003theoretical}, where $\hbar\mathbf{q}$ and $\hbar\omega$ are the momentum transfer and energy loss of the scattered photon, respectively--- have been presented by Dornheim and collaborators~\cite{dornheim_dynamic,dynamic_folgepaper,Dornheim_PRE_2020,Hamann_PRB_2020} based on the stochastic sampling of the dynamic local field correction $G(\mathbf{q},\omega)=-\frac{\mathbf{q}^2}{4\pi} K_\textnormal{xc}(\mathbf{q},\omega)$, see Eq.~(\ref{eq:define_G}) below.
Interestingly, these results have revealed the existence of a non-monotonicity in the dispersion relation of $S(\mathbf{q},\omega)$ at low densities, which phenomenologically resembles the \emph{roton feature} in ultracold quantum liquids~\cite{Trigger,Godfrin2012,Ferre_PRB_2016,Dornheim_SciRep_2022}.
These results have subsequently been explained consistently in terms of the alignment of pairs of electrons~\cite{Dornheim_CommPhys_2022,Dornheim_Force_2022} and as an excitonic mode~\cite{Takada_PRB_2016,Koskelo_PRL_2025}, and extended to the strongly coupled electron liquid regime~\cite{Chuna_PRB_2025,Chuna_JCP_2025}.

In this work, we present an in-depth investigation of the temperature dependence of the analytic continuation at the margin of the strongly coupled electron liquid. This regime is characterized by a density parameter of $r_s=\overline{r}/a_\textnormal{B}=20$ (with $\overline{r}$ and $a_\textnormal{B}$ being the Wigner-Seitz radius and Bohr radius) and $\Theta=k_\textnormal{B}T/E_\textnormal{F}=0.75-8$ (with the temperature $T$ and Fermi energy $E_\textnormal{F}$), indicating a mix of strong coupling, strong thermal excitations and partial quantum degeneracy~\cite{vorberger2025roadmap}. This temperature scan compliments earlier results from \cite{Chuna_PRB_2025,Chuna_JCP_2025}, which conducted scans across density $r_s$. As such, we expect our AC results to be of broad interest across many fields related to quantum many-body theory, with a particular relevance for the study of matter under extreme densities, temperatures and pressures~\cite{vorberger2025roadmap,wdm_book,Dornheim_NatComm_2025,new_POP,Bonitz_POP_2024}.

This work encounters a methodological challenge that is expected of PIMC simulations when scanning across temperature. Essentially, there are two competing effects. Firstly, higher temperatures (i.e., lower inverse temperatures $\beta=1/k_\textnormal{B}T$) lead to a smaller accessible imaginary-time range $\tau\in[0,\beta]$, i.e., less imaginary time data. Secondly, default models such as \emph{effective static approximation} (ESA)~\cite{Dornheim_PRL_2020_ESA,Dornheim_PRB_ESA_2021} become exact in the limit of high temperatures. In particular, we explore the practical aspects of the very recent publication by Chuna \textit{et al.}~\cite{chuna2025noiselesslimitimprovedpriorlimit}, which rigorously demonstrated the higher importance of accurate default models for the analytic continuation compared to the reduction of the Monte Carlo error bars in the PIMC results for $F(\tau)$. These competing effects arise for many path integral methods~\cite{Habershon_JCP_2007, jarrell_bookchapter_2012, rothkopf_FrontiersPhys_2022, buividovich_arXiv_2025} so this makes the methods of this paper of interest outside the UEG community.


The paper is organized as follows: In Sec.~\ref{sec:theory}, we introduce the necessary theoretical background, including a brief discussion of the PIMC method~(\ref{sec:PIMC}), linear-response theory~(\ref{sec:LRT}) and the utilized analytic continuation approaches~(\ref{sec:AC}).
All PIMC simulation results for the ITCF $F(\mathbf{q},\tau)$ and analytic continuation results for the dynamic structure factor $S(\mathbf{q},\omega)$ are presented in Sec.~\ref{sec:results}.
The paper is concluded by a brief discussion and outlook in Sec.~\ref{sec:outlook}.

\section{Theory\label{sec:theory}}

We assume Hartree atomic units (i.e., $m_e = \hbar = k_\textnormal{B} = 1$) throughout the remainder of this work.

\subsection{Path integral Monte Carlo\label{sec:PIMC}}

The basic idea of the PIMC method~\cite{cep} is to manipulate the canonical partition function $Z(N,\beta,\Omega)=\textnormal{Tr}\hat{\rho}$, with $\Omega=L^3$ being the volume of the cubic simulation cell and $\hat\rho=e^{-\beta\hat H}$ the canonical density operator, in coordinate representation.
After a few straightforward manipulations that involve a Trotter decomposition~\cite{Trotter,kleinert2009path}
and the exact semi-group property of $\hat\rho$, we end up with the compact symbolic representation
\begin{eqnarray}\label{eq:partition_function}
    Z(N,\beta,\Omega) = \sumint \textnormal{d}\mathbf{X}\ W(\mathbf{X})\ ,
\end{eqnarray}
where the meta-variable $\mathbf{X}=(\mathbf{R}_0,\dots,\mathbf{R}_{P-1})^T$
contains the coordinates of all $N$ electrons on all $P$ imaginary time slices, with $\mathbf{R}_\alpha=(\mathbf{r}_{\alpha,1},\dots,\mathbf{r}_{\alpha,N})^T$ and $\mathbf{r}_{\alpha,l}$ the coordinate of electron $l$ on imaginary-time slice $\alpha$.
Each such path configuration contributes to the full partition function proportional to the configuration weight $W(\mathbf{X})$, which is an analytically known function that contains both the Ewald interaction between the electrons in the periodic simulation cell~\cite{Fraser_PRB_1996}, and the Gaussian imaginary-time diffusion process~\cite{Dornheim_PTR_2022} that stems from the kinetic energy operator.
We further note that the notation $\sumint\textnormal{d}\mathbf{X}$
takes into account both the integration of all coordinates over the entire volume $\Omega$ as well as the sampling of all possible permutation topologies~\cite{Dornheim_permutation_cycles}, which is required for the correct fermionic anti-symmetry under the exchange of particle coordinates.
In this work, we use the extended ensemble approach~\cite{Dornheim_PRB_nk_2021} as it has been implemented into the publicly available \texttt{ISHTAR} code~\cite{ISHTAR}.

The expectation value of an operator $\hat{A} = \textnormal{Tr}\hat\rho\hat A$ is then computed as
\begin{eqnarray}\label{eq:expectation_value}
    \braket{\hat A} 
 \nonumber   &=& \frac{1}{Z(N,\beta,\Omega)}\sumint\textnormal{d}\mathbf{X}\ W(\mathbf{X}) A(\mathbf{X}) \\\label{eq:approx}
    &\approx& \frac{1}{N_\textnormal{MC}}\sum_{l=1}^{N_\textnormal{MC}} A(\mathbf{X}_l)\ , \nonumber
\end{eqnarray}
where the $N_\textnormal{MC}$ random configurations $\mathbf{X}_l$ in the last line are distributed according to $P(\mathbf{X})=W(\mathbf{X})/Z(N,\beta,\Omega)$.
In the context of the present work, the key observable is given by the imaginary-time density--density correlation function 
\begin{eqnarray}\label{eq:define_F}
    F(\mathbf{q},\tau) = \braket{\hat{n}(\mathbf{q},0)\hat{n}(-\mathbf{q},\tau)}\ ,
\end{eqnarray}
which is obtained by correlating two electronic densities (in reciprocal space) at an imaginary-time distance $\tau$, and which is connected to the dynamic structure factor (DSF) $S(\mathbf{q},\omega)$ via Eq.~(\ref{eq:ACproblem}).
For completeness, we note that $F(\mathbf{q},\tau)$ contains a wealth of physical information~\cite{Dornheim_MRE_2023,Dornheim_PTR_2022,Dornheim_PRB_2023,Dornheim_NatComm_2022}, including its straightforward relation to the static linear density response function via the imaginary-time version of the fluctuation--dissipation theorem~\cite{Dornheim_MRE_2023}, see Eq.~(\ref{eq:static_chi}) below.

An additional practical obstacle is given by the fermionic antisymmetry under particle exchange, which means that the configuration weight $W(\mathbf{X})$ can be either positive or negative.
The required fermionic re-weighting leads to a cancellation of positive and negative contributions to Eqs.~(\ref{eq:partition_function}) and (\ref{eq:expectation_value}), leading to a vanishing signal-to-noise ratio in the limit of low temperatures and large numbers of particles.
The resulting exponential computational bottleneck is known as the \emph{fermion sign problem}~\cite{troyer,dornheim_sign_problem},
and constitutes one of the most stifling bottlenecks across quantum many-body theory.
While many strategies to mitigate the sign problem have been proposed over the years, e.g., Refs.~\cite{Ceperley1991,Blunt_PRB_2014,review,Dornheim_POP_2017,Egger_PRE_2000,Chin_PRE_2015,Dornheim_NJP_2015,Schoof_CPP_2015,Yilmaz_JCP_2020,Hirshberg_JCP_2020,Dornheim_JCP_2020,Joonho_JCP_2021,Xiong_JCP_2022,Dornheim_JCP_2023,Xiong_PRE_2023,Xiong_JCP_2025,dornheim2025taylorseriesperspectiveab}, currently no universal solution that is free of particular limitations or approximations exists.
In the current work, we carry out direct PIMC simulations, which are afflicted with the full sign problem, and our results are exact within the given Monte Carlo error bars.
A convenient measure for the amount of cancellations within the simulations is given by the so-called \emph{average sign} $S$~\cite{dornheim_sign_problem}, and we find
$S=0.3225(3)$ ($\Theta=0.75$) and $S=0.9895(1)$ ($\Theta=8$) at the lower and higher ends of the considered range of temperatures for $N=34$ electrons at $r_s=20$; this means that full simulations are feasible with manageable computation effort.

All PIMC raw data for $F(\mathbf{q},\tau)$ are available in an online repository~\cite{repo}.

\subsection{Linear response theory\label{sec:LRT}}

Linear response theory is one of the most widely used tools of physics. 
Within warm dense matter theory~\cite{Dornheim_review,vorberger2025roadmap,wdm_book,Bonitz_POP_2024}, it is used, e.g., to estimate stopping power~\cite{Moldabekov_PRE_2020}, effective potentials~\cite{Dornheim_Force_2022}, and ionization potential depression~\cite{Zan_PRE_2021}.
For this work, the key quantity of interest is the dynamic density response function $\chi(\mathbf{q},\omega)$, which describes the response to an external harmonic perturbation of wavevector $\mathbf{q}$ and frequency $\omega$ in the limit of an infinitesimal perturbation amplitude $A$.
It is often expressed as~\cite{kugler1}
\begin{eqnarray}\label{eq:define_G}
    \chi(\mathbf{q},\omega) = \frac{\chi_0(\mathbf{q},\omega)}{1 - \frac{4\pi}{\mathbf{q}^2}\left[1-G(\mathbf{q},\omega)\right]\chi_0(\mathbf{q},\omega)}\ ,
\end{eqnarray}
with $\chi_0(\mathbf{q},\omega)$ being the (temperature-dependent) Lindhard function describing the known density response of the non-interacting ideal Fermi gas~\cite{giuliani2008quantum}.
The complete wavenumber and frequency resolved information about dynamic XC-effects is encoded into the aforementioned dynamic local field correction $G(\mathbf{q},\omega)$, and setting $G(\mathbf{q},\omega)\equiv0$ in Eq.~(\ref{eq:define_G}) corresponds to the ubiquitous \emph{random phase approximation} (RPA) that describes the density response on the mean-field level~\cite{pines}.
The local field correction is very important in its own right; it constitutes important input for a great variety of applications~\cite{transfer1,transfer2,Zan_PRE_2021,Dornheim_PRL_2020_ESA,pribram,Patrick_JCP_2015,Fortmann_PRE_2010,bespalov2026experimentalevidencebreakdownuniformelectrongas,ceperley_potential,Kukkonen_PRB_2021,Poole_PRR_2024}, and also constitutes the central object for the closure relations within dielectric theory~\cite{vs_original,stls_original,stls,schweng,stolzmann,tanaka_hnc,stls2,Tolias_JCP_2021,Tolias_JCP_2023,Tolias_PRB_2024,Tolias_CPP_2025,kalkavouras2026dielectricformalism2duniform}.

In this work, the main application of linear response theory is creating improved default models of the DSF for the AC algorithms. The DSF is related to the dynamic density response Eq.~(\ref{eq:define_G}) by the fluctuation--dissipation theorem~\cite{giuliani2008quantum}
\begin{eqnarray}\label{eq:FDT}
S(\mathbf{q},\omega) = - \frac{\textnormal{Im}\chi(\mathbf{q},\omega)}{\pi n (1-e^{-\beta\omega})}\ ,
\end{eqnarray}
For comparison purposes, we compute the DSF in the RPA, (\textit{i.e.}, $G_\textnormal{RPA}(\mathbf{q},\omega)\equiv 0$) and for the AC algorithms' default model we only use the \emph{static approximation}~\cite{dornheim_dynamic} $G_\textnormal{static}(\mathbf{q},\omega)\equiv G(\mathbf{q},0)$. We obtain the exact static local field correction from the imaginary-time generalization of Eq.~(\ref{eq:FDT})~\cite{Dornheim_MRE_2023,bowen2},
\begin{eqnarray}\label{eq:static_chi}
    \chi(\mathbf{q},0) = - \frac{N}{\Omega}\int_0^\beta\textnormal{d}\tau\ F(\mathbf{q},\tau)\ ,
\end{eqnarray}
and then solving Eq.~(\ref{eq:define_G}) for $G(\mathbf{q},0)$.

\subsection{Analytic continuation\label{sec:AC}}
In this section, we discuss the analytic continuation to compute the DSF $S(\mathbf{q},\omega)$ from the imaginary-time density--density correlation function $F(\mathbf{q},\tau)$.
For these observables, the general kernel equation (\ref{eq:kernel}) becomes a two-sided Laplace transform,
\begin{eqnarray}\label{eq:ACproblem}
F(\mathbf{q},\tau) = \int_{-\infty}^\infty \textnormal{d}\omega\ S(\mathbf{q},\omega)\ e^{-\tau\omega} \, ;
\end{eqnarray}
its inversion is a well-known, exponentially ill-posed inverse problem~\cite{Jarrell_PhysRep_1996,Chuna_JPA_2025}.

Over the last few decades, many analytic continuation methods have been created, e.g., Refs.~\cite{MEM_original,Jarrell_PhysRep_1996,Mishchenko_PRB_2000,Vitali_PRB_2010,Sandvik_PRE_2016,Bertaina_GIFT_2017,dornheim_dynamic,dynamic_folgepaper,Otsuki_PRE_2017,Otsuki_JPSJ_2020,SHAO20231,Filinov_PRB_2023,Chuna_JPA_2025,BENEDIXROBLES2026109904}. The methods typically estimate $S$ by solving an optimization of the following form
\begin{align}\label{eq:typical_optimization_problem}
    S(\omega) = \min_x \quad  - \frac{1}{2} \Vert A x - b \Vert + \lambda r[ \cdot ] \, .
\end{align} 
The fidelity term $\frac{1}{2} \Vert A x - b \Vert$ is common among approaches and here we use the sum of squared errors (SSE) goodness of fit metric. The most prevalent approaches (stochastic and regularized optimization) differ in whether they include the regularization term $r$ in Eq.~(\ref{eq:typical_optimization_problem}). Authors uniformly recommend that all approaches average over many representative solutions~\cite{Han_PRB_2022, Goulko_PRB_2017, gunnarsson_PRB-MEM_2010} obtained from the discrete formulation of Eq.~(\ref{eq:ACproblem}), but differ on how the representative solutions are obtained. For this work, we use \eqref{eq:typical_optimization_problem} with the most popular regularization, the Shannon-Jaynes information entropy 
\begin{align}\label{eq:SJEntropy}
    r[\cdot] \rightarrow S_{SJ}[x | \mu] = \sum_k S(\omega_k) - \mu(\omega_k) - S(\omega_k) \log \left(\frac{S(\omega_k)}{\mu(\omega_k)} \right) \, ,
\end{align}
where $\mu(\omega)$ is the default model. We collect our set of representative solutions via leave-one-out resampling of the data, as has been discussed in detail elsewhere~\cite{Chuna_PRB_2025}.

Recently, a kernel-based reformulation of Eq.~(\ref{eq:typical_optimization_problem}) has been presented by Benedix-Robles \textit{et al.}~\cite{BENEDIXROBLES2026109904}, which represents the DSF $S(\omega)$ as a linear combination of parameterized functions,
\begin{align}\label{eq:kernel_parameterization}
    S(\omega) = \sum_{j=1}^{N_c} c_j K_j(\omega) \, ,
\end{align}
where we suppressed the wavevector $\mathbf{q}$ for simplicity.
In this work we assume Gaussian kernels with a center $\mu_j$ and a width $\sigma_j$ that define $K_j(\omega) = \exp[-(\omega - \mu_j)^2/\sigma^2_j]$, but other kernels could, in principle, be constructed. Eq.~(\ref{eq:kernel_parameterization}) maps the optimization problem from an $N_\omega$-dimensional space to a smaller, $N_c$-dimensional space. Substituting Eq.~(\ref{eq:kernel_parameterization}) into (\ref{eq:ACproblem}) produces
\begin{subequations}
\begin{align} \label{eq:reformulatedAC}
    F(\tau) &= \sum_{j=1}^{N_c} c_j R_j(\tau) \, ,
    \\ R_j(\tau) &= \int_{-\infty}^\infty  e^{- \tau \omega} K_j(\omega)\ \textnormal{d}\omega \ .\label{eq:kerneltransform}
\end{align}
\end{subequations}
By discretizing $\tau$, the Laplace transformed kernel $R_j(\tau)$ becomes a matrix, leading to an optimization over the kernel coefficients,
\begin{align}\label{eq:PyLITcostfunction}
    S(\omega) = \min_c \left\{ \Big\Vert F - R c \Big\Vert^2 + \lambda r(\cdot) \right\} \, ,
\end{align}
where $c \in \mathbb{R}^{N_c}$ and $N_c$ is the number of kernels. Eq.~(\ref{eq:PyLITcostfunction}) is the kernel-based formulation of (\ref{eq:typical_optimization_problem}) and there are many different kernel choices that can be made. In the case of Dirac delta kernels located on a fixed $\omega$ grid, Eq.~(\ref{eq:PyLITcostfunction}) recovers the original formulation Eq.~(\ref{eq:typical_optimization_problem}); there is no dimensional reduction in this case. 

In this work, we solve both the typical formulation \eqref{eq:typical_optimization_problem} and the kernel formulation \eqref{eq:PyLITcostfunction} regularized by the Shannon-Jaynes information entropy \eqref{eq:SJEntropy}. For both formulations, we determine the regularization weight $\lambda$ via the so-called $\chi^2$-kink algorithm~\cite{Kaufmann_CPC-anacont_2023}, which selects the largest regularization weight such that the deviations in the fidelity term are small. For the kernel formulation, we optimize Eq.~(\ref{eq:PyLITcostfunction}) using the publicly available \texttt{PyLIT} code~\cite{BENEDIXROBLES2026109904} with Gaussian kernels $j= 0,\ldots, 49$. \texttt{PyLIT} determines the parameters of these Gaussians by fitting to the default model $\mu(\omega)$. For the typical formulation, we optimize \eqref{eq:typical_optimization_problem} using Bryan's MEM \cite{bryan_EuroBiophys_1990, Jarrell_PhysRep_1996, Asakawa_PPNP_2001}, as implemented by Chuna et al.~\cite{Chuna_JPA_2025} and freely available online~\cite{github_MEMcode}. We solve on a sparse $\omega$ grid, $N_\omega \sim \mathcal{O}[50]$, so that the dimension of the traditional optimization problem \eqref{eq:typical_optimization_problem} is reflective of the dimensionality of the \texttt{PyLIT} approach, and any compression that comes from the singular value decomposition in Bryan's algorithm is an added bonus. The sparse $\omega$-grid is acceptable because we expect the DSF of the UEG to have a smooth structure at the present conditions~\cite{dornheim_dynamic,Chuna_PRB_2025,Chuna_JCP_2025,Filinov_PRB_2023}. We fill in the dense $\omega$-grid points by interpolating the MEM estimate, this is effectively what is done whenever a MEM solution is plotted. 



\section{Results\label{sec:results}}

All results have been computed for the unpolarized UEG, i.e., with an equal number of spin-up and spin-down electrons $N^\uparrow=N^\downarrow=N/2$ for $N=34$. For $r_s = 20$ and at $\Theta=0.75$, there are 1000 independent MCMC seeds, at $\Theta=1, \, 2$ there are $280$ seeds, at $\Theta=4, \,8$ there are $277$ seeds. To compute the data, we conduct leave-one-out binning across the seeds and for all the data, the variance of the mean is $\delta F \approx 10^{-3}-10^{-4}$. The error estimate is computed for each leave-one-out-bin via Hatano's error formula~\cite{hatano1994data} and verified using leave-one-out binning~\cite{berg_book_2004}. An online repository with our PIMC results for $F(\mathbf{q},\tau)$ is available in Ref.~\cite{repo}.

\begin{figure*}[t]
    \centering
    \includegraphics[width=.5\linewidth]{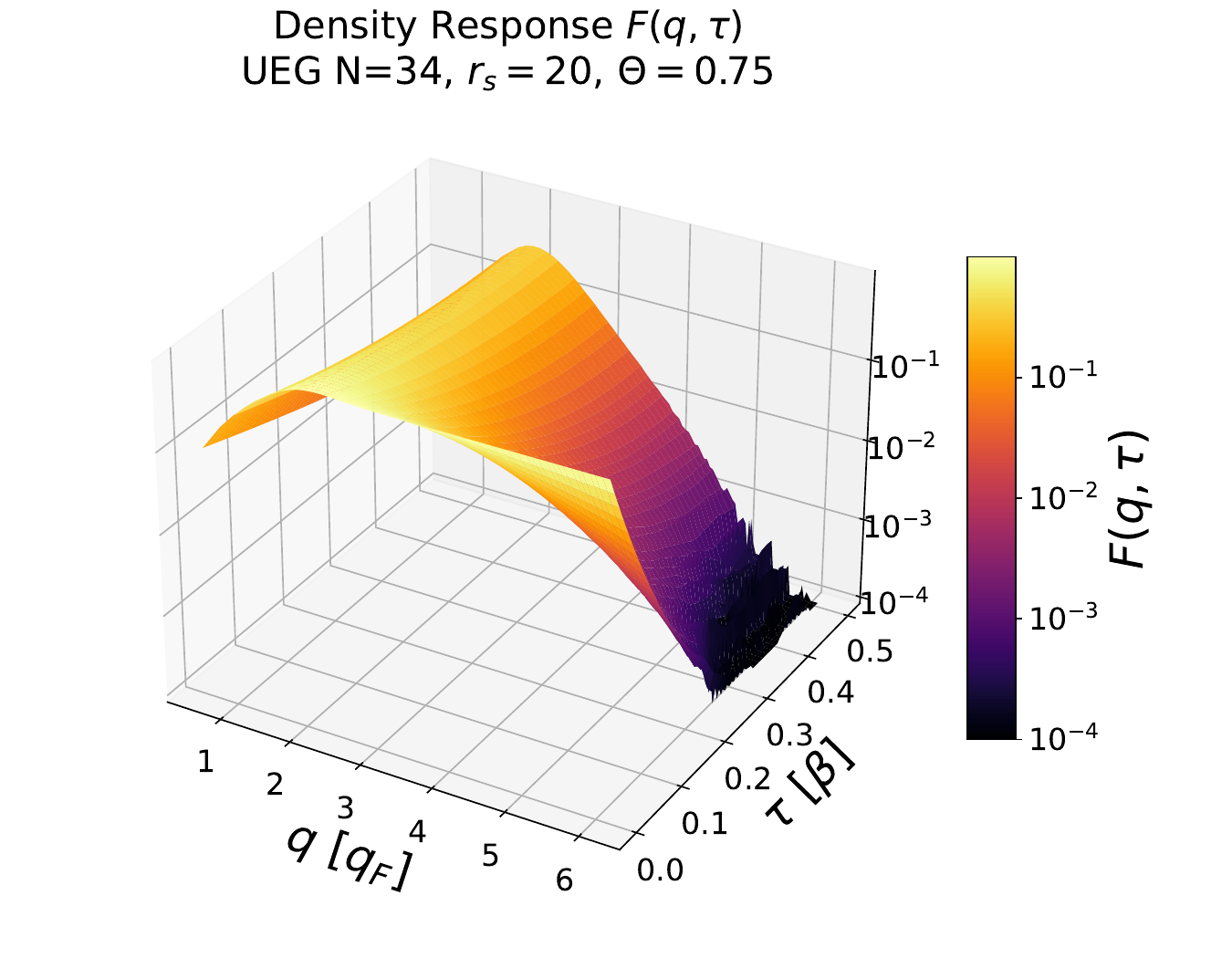}%
    \includegraphics[width=.5\linewidth]{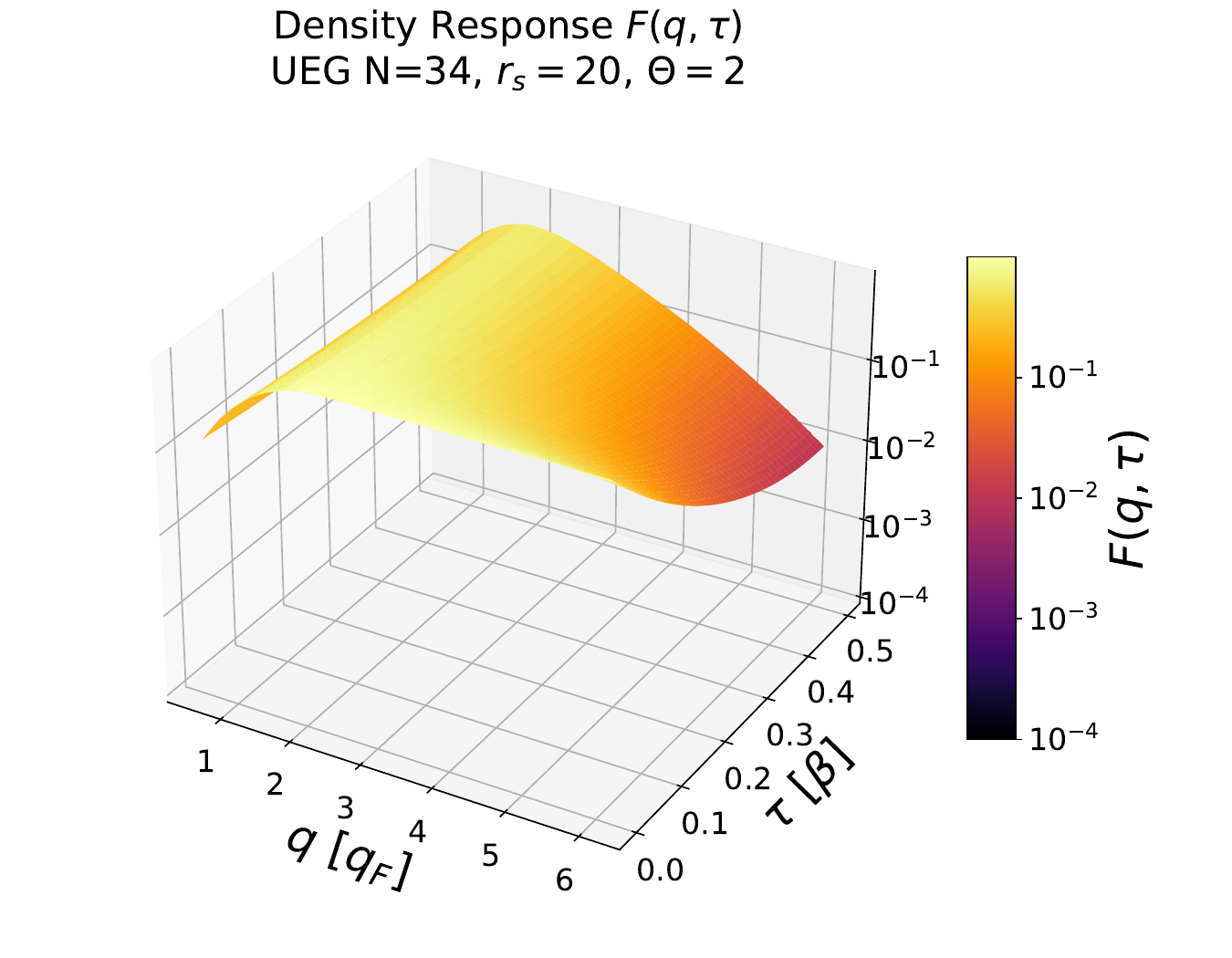}
    \caption{\emph{Ab initio} PIMC results for the ITCF $F(\mathbf{q},\tau)$ of the UEG with $N=34$ at $r_s=20$ for $\Theta=0.75$ (left) and $\Theta=2$ (right).}
    \label{fig:3D}
\end{figure*}
Let us start our analysis with a brief discussion of our PIMC estimates of the imaginary-time density--density correlation function $F(\mathbf{q},\tau)$ [cf.~Eq.~(\ref{eq:define_F}) above], which we show in Fig.~\ref{fig:3D} for $r_s=20$ for $\Theta=0.75$ (left panel) and $\Theta=2$ (right panel) in the $q$-$\tau$-plane.
We note that it is sufficient to restrict ourselves here to the interval of $0\leq\tau\leq\beta/2$ due to the symmetry relation
\begin{eqnarray}\label{eq:ITCF_symmetry}
    F(\mathbf{q},\tau) = F(\mathbf{q},\beta-\tau)\ .
\end{eqnarray}
Eq.~(\ref{eq:ITCF_symmetry}) directly follows from the detailed balance relation~\cite{Dornheim_NatComm_2022},
\begin{eqnarray}
    S(\mathbf{q},-\omega) = S(\mathbf{q},\omega)\ e^{-\beta\omega}\ ,
\end{eqnarray}
and has recently been employed for the model-free temperature diagnostics of XRTS measurements at extreme conditions~\cite{Dornheim_NatComm_2022,Dornheim_T2_2022}, see also the discussion in Sec.~\ref{sec:outlook} below.
The \emph{static} limit of $\tau\to0$ of the ITCF is given by the static structure factor $F(\mathbf{q},0)=S(\mathbf{q})$, which constitutes the normalization of $S(\mathbf{q},\omega)$ and is related to the electronic pair correlation function via a straightforward Fourier transform~\cite{hansen2013theory}.
For the UEG, we have $\lim_{q\to0}S(\mathbf{q})\sim \mathbf{q}^2$ (a consequence of perfect screening~\cite{kugler_bounds}) and also the generally applicable $\lim_{q\to\infty}S(\mathbf{q})=1$ (a consequence of the perfect correlation of each electron with itself, and the lack of any correlations with other electrons in this limit).
For a given fixed wavevector $\mathbf{q}$, the ITCF monotonically decays with $\tau$ (up to $\tau=\beta/2$, beyond which it symmetrically increases again); this decay is mainly a consequence of the decay of self correlations as it has been discussed in detail by Dornheim \textit{et al.}~\cite{Dornheim_PTR_2022}, although other effects such as electronic correlations~\cite{Dornheim_MRE_2023,Chuna_JCP_2025} and electron-ion correlations~\cite{Dornheim_MRE_2024} can influence the decay rate.

This decay becomes increasingly steep for increasing wavenumbers $q$, which can be understood in different ways.
First, we note that increasing $q$ correspond to small wavelengths $\lambda = 2\pi / q$, and the imaginary-time diffusion process has a stronger impact on smaller length scales.
Second, it is easy to see that the derivatives of $F(\mathbf{q},\tau)$ with respect to $\tau$ around $\tau=0$ are directly related to the frequency moments of $S(\mathbf{q},\omega)$; see Ref.~\cite{Dornheim_PRB_2023} for a detailed discussion.
The first frequency moment of the dynamic structure factor is governed by the universal frequency sum rule~\cite{chuna2026merminsdielectricfunctionfsum}, which means that the slope of $F(\mathbf{q},\tau)$ around the origin scales as $\sim q^2$~\cite{Dornheim_MRE_2023,Dornheim_SciRep_2024}. This is a direct consequence of quantum delocalization and also explains the increasingly steep decay of $F(\mathbf{q},\tau)$ for large wavenumbers.
Importantly, lower temperatures automatically correspond to a larger $\tau$-interval on which $F(\mathbf{q},\tau)$ is being resolved.
This, in turn, means that the ITCF contains less information in the limit of high temperatures.
This becomes immediately evident from Fig.~\ref{fig:3D}, where we observe a pronounced structure for $\Theta=0.75$ and a relatively flat profile even for $\Theta=2$.

\begin{figure*}
    \centering
    \includegraphics[width=\linewidth]{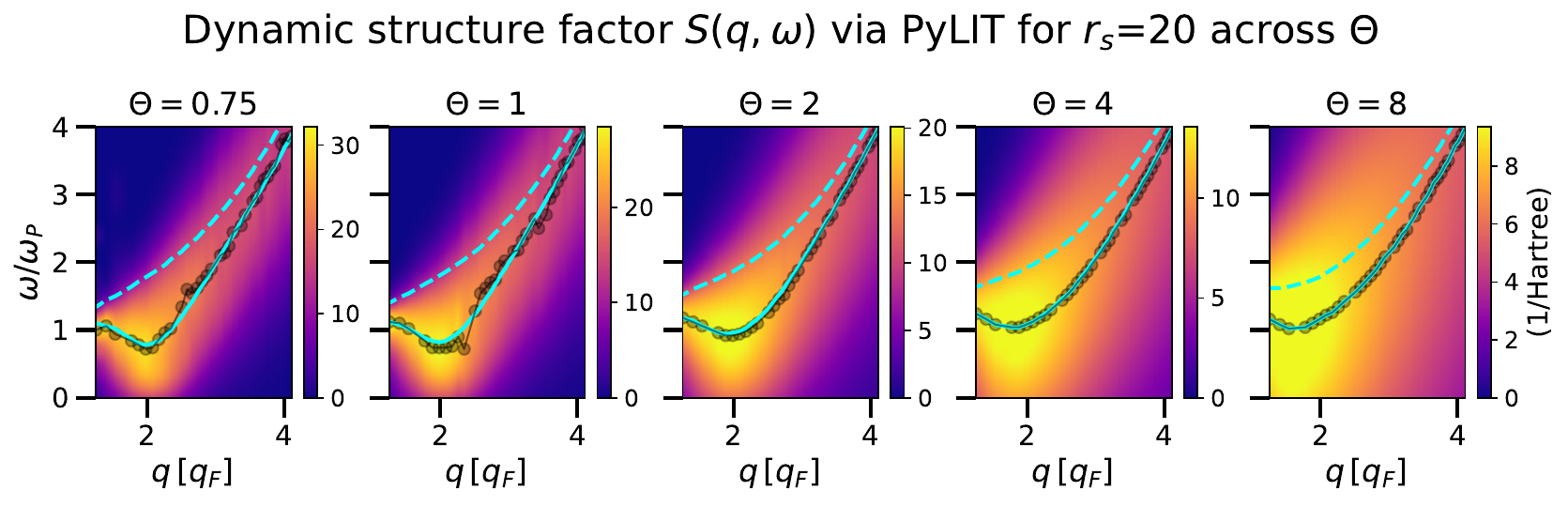}
    \includegraphics[width=\linewidth]{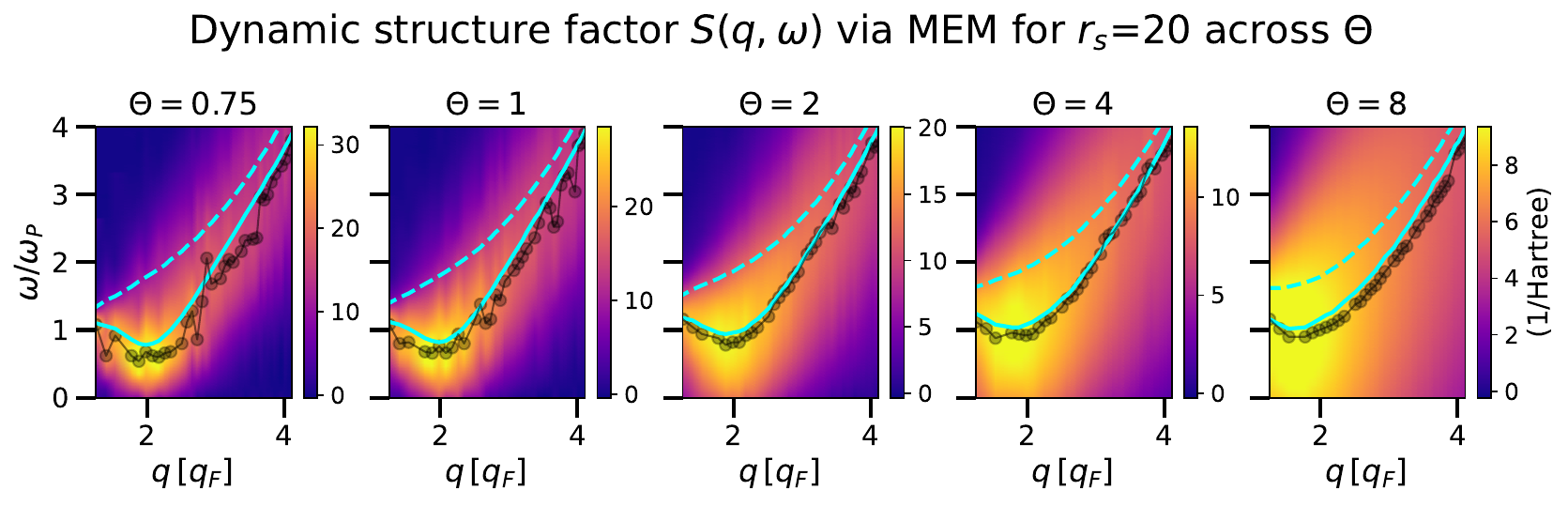}
    \caption{Heatmaps: analytic continuation results for the dynamic structure factor of the UEG with $N=34$ at $r_s=20$ for different values of the reduced temperature $\Theta$; frequencies are given in units of the plasma frequency $\omega_\textnormal{p}=\sqrt{3/r_s^3}$ and wavenumbers in units of the Fermi wavenumber $q_\textnormal{F}=(9\pi/4)^{1/3}/r_s$. The top and bottom rows have been obtained using \texttt{PyLIT} and the standard MEM. We note that the color scale changes between plots.
    The top and bottom cyan curves show the position of the maximum of $S(\mathbf{q},\omega)$ computed within RPA and \emph{static approximation} [i.e., $G(\mathbf{q},\omega)\equiv0$ and $G(\mathbf{q},\omega)\equiv G(\mathbf{q},0)$ in Eq.~(\ref{eq:define_G})], and the black curves show the same information for the analytic continuation. 
    }
    \label{fig:AC}
\end{figure*}
Let us next come to the main topic of this investigation, which is the analytic continuation to reconstruct the dynamic structure factor $S(\mathbf{q},\omega)$ from our PIMC results for the ITCF $F(\mathbf{q},\tau)$.  In Fig.~\ref{fig:AC}, we show the reconstructed dynamic structure factors as heatmaps, with the top and bottom rows showing \texttt{PyLIT} and MEM results, respectively, and the different columns corresponding to the five considered values of the reduced temperature $\Theta$.
In addition, we have included the position of the maximum in the dynamic structure factor with respect to the frequency $\omega$ estimated within the RPA and within the \emph{static approximation} as the top and bottom cyan curves, and the same quantity estimated from the full analytic continuation as the black curves.
As a general trend, we find that the RPA substantially overestimates the true position of the maximum of $S(\mathbf{q},\omega)$ compared to all other results.
This is consistent with previous investigations at similar conditions~\cite{dornheim_dynamic,Chuna_PRB_2025,Chuna_JCP_2025,Dornheim_CommPhys_2022}, and it is readily explained by the lack of XC-effects within the RPA, which are the key mechanism behind the electronic pair alignment model that has been devised to explain the roton-type feature at intermediate wavenumbers $q\sim 2q_\textnormal{F}$.
Interestingly, the roton minimum persists even for the highest considered temperature of $\theta=8$, although with a substantially reduced depth.

Focusing next on the analytic continuation results. The main difference between the two methods is that the dispersion relation from \texttt{PyLIT} approach very closely follows the \emph{static approximation} prior, while the standard MEM dispersion relation has a deeper roton minimum. We recall that while both methods use the \emph{static approximation} as the default model $\mu(\omega)$, \texttt{PyLIT} also constructs its Gaussian kernel basis set to represent $\mu(\omega)$ (see Sec.~\ref{sec:AC} above) which introduce additional regularization. The MEM is more consistent with arguably the most accurate analytic continuation results for the UEG presented in Refs.~\cite{dornheim_dynamic,dynamic_folgepaper}. At the same time, the MEM DSF results are less stable, having more fluctuations between adjacent wavenumbers. This leads to a high variation in its dispersion relation. These results are expected since the typical trade-off for analytic continuation solvers is between between a stable bias and unstable fidelity and they highlight the value of comparing alternative approaches.

\begin{figure*}
    \centering
    \includegraphics[width=\linewidth]{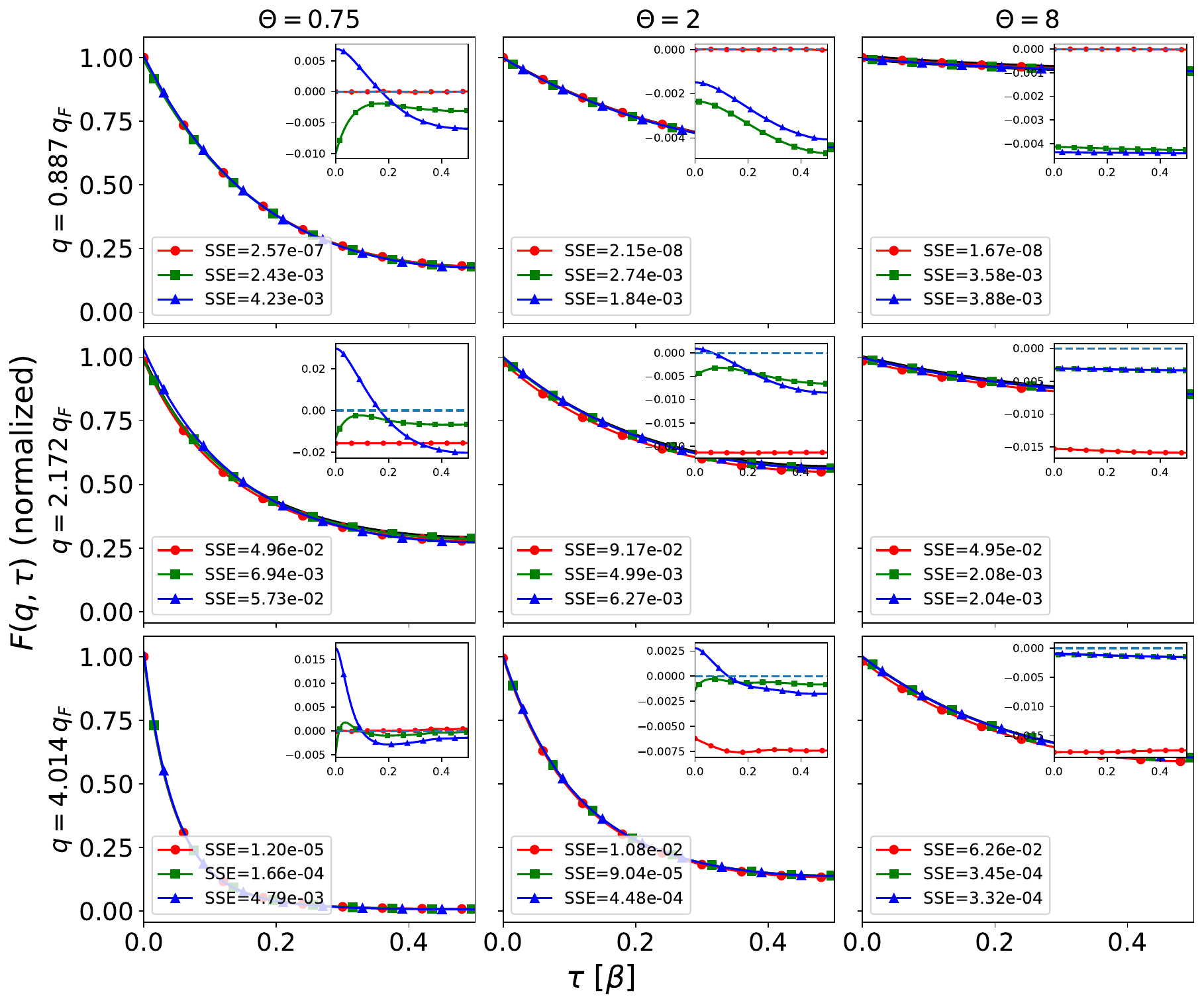}
    \caption{ITCF $F(\mathbf{q},\tau)$ for selected wavenumbers $q$ at $\Theta=0.75$ (left), $\Theta=2$ (middle) and $\Theta=8$ (right). The red symbols correspond to the results for typical formulation with Bryan's MEM algorithm, the results for kernel formulation with PyLIT code are green, and the static approximation is depicted in blue. The insets show the absolute deviation from the PIMC ground truth with the same color code.}
    \label{fig:Ftau_with_residual}
\end{figure*}
Next we compare the Laplace transformed DSFs to the ITCF $F(\mathbf{q},\tau)$ for selected wavenumbers and temperatures in Fig.~\ref{fig:Ftau_with_residual}. Overall, hardly any deviations between the red, green and blue data sets that show results from standard MEM, \texttt{PyLIT} and the \emph{static approximation}, respectively, can be resolved with the naked eye for any of the depicted cases. The insets show the absolute deviation of these data from the PIMC results with the same color code. We note the high quality of the \emph{static approximation}, which exhibits its largest deviations to the PIMC ground truth for intermediate wavenumbers where dynamic XC-effects are expected to be most important~\cite{dornheim_dynamic}. This is important in the context of Chuna et al. which established that a high quality default model is sufficient to guarantee the validity of the MEM and Bryan's algorithm. 

Focusing the MEM and PyLIT results in Fig.~\ref{fig:Ftau_with_residual}, we consider the SSE, defined in Eq.(\ref{eq:typical_optimization_problem}), to assess the quality of the analytic continuation. We see that the standard MEM approach leads to a very good agreement with PIMC for the smallest wavenumber, where the dynamic structure factor is relatively narrow. This agreement deteriorates at larger wavenumber, where the interpolation algorithm must interpolate over larger $\omega$ intervals. The deterioration occurs because interpolating solutions obtained on a sparse $\omega$-grid and to a dense $\omega$-grid means that Bryan's MEM optimized a different goodness-of-fit metric than what is being presented. We note that the MEM always achieves SSE $< 10^{-5}$, with typical values $\approx 10^{-7}$ on its sparse grid, it is only the interpolated DSF has a larger SSE. For the \texttt{PyLIT} results, we see they generally show a similar level of agreement compared to the \emph{static approximation}, which that indicates using the default model to tune the kernel set and regularize the optimization introduces significant bias and reduces the fidelity of the solution. Additionally, for the kernel formulation, the PyLIT code optimizes the kernel coefficients with fixed kernel parameters, it does not optimize the goodness-of-fit of the DSF on the dense grid. So neither method is being compared on the metric it optimized, but since they do not share an optimization problem this chosen metric is a natural one.

\section{Discussion and Outlook}\label{sec:outlook}

In this work, we have explored the temperature dependence of the dynamic structure factor $S(\mathbf{q},\omega)$ by computing the analytic continuation of quasi-exact PIMC estimates of the imaginary-time density--density correlation function $F(\mathbf{q},\tau)$.
To this end, we have employed two related but distinct approaches: uniform discretization as it has been implemented in the standard MEM and a kernel reformulation as it has been implemented by the \texttt{PyLIT}
code~\cite{BENEDIXROBLES2026109904}.
By making the dimensions of the two problems similar and using the same regularization (\textit{i.e.}, Shannon-Jaynes entropy with \emph{static approximation} prior), We have shown that the non-linear Gaussian kernels lead to a higher stability, but this comes at the cost of introducing substantial bias towards the default model. While the SSE are small, the associated $\chi^2$ exceeds the expected threshold $\chi^2 > 2$. Thus the results are inconclusive. This paper demonstrates the need to optimize the Gaussian parameters to represent the data, rather than to represent the default model. By doing so, this should allow the PyLIT code to achieve a better $\chi^2$ value.

Future work might focus on the application of analytic continuation approaches to PIMC results for the ITCF of real two-component systems, such 
as warm dense hydrogen~\cite{Dornheim_MRE_2024,Dornheim_JCP_2024} and beryllium~\cite{Dornheim_JCP_2024,Dornheim_NatComm_2022,Dornheim_POP_2025,schwalbe2025staticlineardensityresponse}, and on the reconstruction of other properties such as the single-particle spectral function $A(\mathbf{q},\omega)$~\cite{hamann2026abinitiopathintegralmonte}
and the heat conductivity~\cite{efremkin2026computationthermalconductivitybased}.

The study of realistic systems and experimental data would benefit from the parameter-free estimates that AC can provide and these estimates would be greatly supported by directly extracting physics information from $F(\mathbf{q},\tau)$. This can be done in the following ways: by using the relation of the ITCF to the frequency moments of the dynamic structure factor~\cite{Dornheim_PRB_2023,Dornheim_MRE_2023}, by analyzing its $\tau$-decay~\cite{Dornheim_MRE_2023,Chuna_JCP_2025}, by forward-fitting XRTS models to extract chemical model parameters such as the ionization degree and ionization potential depression~\cite{Bellenbaum_PRR_2025}.
These efforts might be further complemented by the extraction of  the generalized dynamic Matsubara density response~\cite{Tolias_JCP_2024,Dornheim_PRB_2024,Dornheim_EPL_2024,MOLDABEKOV2025104144,Dornheim_CPP_2025} for a variety of systems.
While directly extracting physics information from $F(\mathbf{q},\tau)$ is important in its own right, such efforts can inspire the development of new methodologies, such as the very recent model-free ITCF approach for the interpretation of XRTS experiments~\cite{Dornheim_NatComm_2022,Dornheim_T2_2022,Dornheim_SciRep_2024,Dornheim_POP_2025,schwalbe2025staticlineardensityresponse}, which has already been
applied to experimental measurements at the National Ignition Facility in Livermore~\cite{Dornheim_SciRep_2024,boehme2023evidence,Vorberger_PLA_2024,Dornheim_NatComm_2025,Dornheim_POP_2025,schwalbe2025staticlineardensityresponse},
the Linac Coherent Light Source (LCLS) in Stanford~\cite{Dornheim_NatComm_2022,Dornheim_T2_2022,Bellenbaum_APL_2025},
the Omega Laser facility in Rochester~\cite{Dornheim_NatComm_2022,Dornheim_T2_2022,Schoerner_PRE_2023}, the European XFEL in Germany~\cite{Dornheim_SciRep_2024,Smid_SciRep_2026}, as well as the Shengguan-II laser facility in China~\cite{shi2025firstprinciplesanalysiswarmdense}.

\begin{acknowledgements}
Tobias Dornheim gratefully acknowledges funding from the Deutsche Forschungsgemeinschaft (DFG) via project DO 2670/1-1.
This work has received funding from the European Union's Just Transition Fund (JTF) within the project \emph{R\"ontgenlaser-Optimierung der Laserfusion} (ROLF), contract number 5086999001, co-financed by the Saxon state government out of the State budget approved by the Saxon State Parliament. This work has received funding from the European Research Council (ERC) under the European Union’s Horizon 2022 research and innovation programme (Grant agreement No. 101076233, "PREXTREME"). 
Views and opinions expressed are however those of the authors only and do not necessarily reflect those of the European Union or the European Research Council Executive Agency. Neither the European Union nor the granting authority can be held responsible for them.
Maximilian P.~B\"ohme's work was performed under the auspices of the U.S. Department of Energy by Lawrence Livermore National Laboratory under Contract No.~DE-AC52-07NA27344. M.P.B. was supported by Laboratory Directed Research and Development (LDRD) Grant No.~25-ERD-047.
Computations were performed on a Bull Cluster at the Center for Information Services and High-Performance Computing (ZIH) at Technische Universit\"at Dresden, at the Norddeutscher Verbund f\"ur Hoch- und H\"ochstleistungsrechnen (HLRN) under grant mvp00024. Futher computing support for this work came from the Lawrence Livermore National Laboratory (LLNL) Institutional Computing Grand Challenge program.
\end{acknowledgements}

\bibliography{bibliography}

\end{document}